\documentclass[twocolumn,aps,prl,10pt,showpacs,floatfix]{revtex4-1}
\usepackage{bm}
\usepackage{amsfonts}
\usepackage{amssymb}
\usepackage{amsmath}
\usepackage{graphicx}

\begin{document}

\title{Trapping and guiding bodies by gravitational waves endowed with angular momentum}
\author{Iwo Bialynicki-Birula}\email{birula@cft.edu.pl}
\affiliation{Center for Theoretical Physics, Polish Academy of Sciences\\
Aleja Lotnik\'ow 32/46, 02-668 Warsaw, Poland}
\author{Szymon Charzy{\' n}ski}\email{szycha@fuw.edu.pl}
\affiliation{Faculty of Physics, University of Warsaw, Pasteura 5, 02-093 Warsaw, Poland}
\begin{abstract}
Trapping of bodies by waves is extended from electromagnetism to gravity. It is shown that gravitational waves endowed with angular momentum may accumulate near its axis all kinds of cosmic debris. The trapping mechanism in both cases can be traced to the Coriolis force associated with the local rotation of the space metric. The same mechanism causes the Trojan asteroids to librate around the Sun-Jupiter stable Lagrange points $L_4$ and $L_5$. Trapping of bodies in the vicinity of the wave center could also be related to the formation of galactic jets.
\end{abstract}
\pacs{4.20.-q,4.30.-w,4.70.-s}
\maketitle

\begin{figure}
\includegraphics[width=0.4\textwidth]{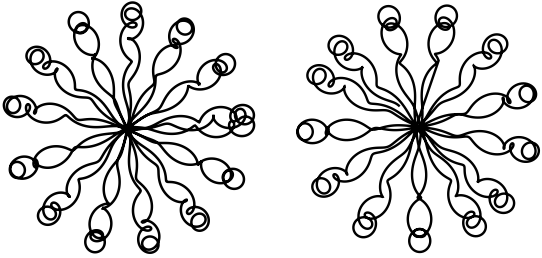}
\vspace{0.3cm}
\caption{Projection of the trajectory on the $xy$ plane for a charged particle trapped by the electromagnetic wave (left) and for a massive particle trapped by the gravitational wave (right). The parameters in the gravitational case were chosen to show the correspondence with the electromagnetic case and not to describe a realistic situation.}\label{fig1}
\end{figure}

It has been established in Refs.~\cite{bessel1,bessel2,bessel3} that electromagnetic waves endowed with angular momentum (for example, Bessel beams) can trap charged particles in the vicinity of the beam center. In this Letter we prove that the gravitational waves carrying angular momentum have the same property. In Fig.\ref{fig1} we plotted the projection of the particle orbit onto the plane perpendicular to the beam direction for the electromagnetic and the gravitational case. The close similarity between the two plots is the best proof that in both case we are dealing with very similar phenomena.
The mechanism which is responsible for trapping of bodies near the center of the gravitational wave is the same as that encountered in many diverse physical systems: it is the Coriolis force. This force leads to stable orbits of Trojan asteroids \cite{danby} and ions in the Paul trap \cite{paul}, to stable wave packets of electrons in high Rydberg states moving in a circularly polarized electromagnetic wave \cite{lagr} and electrons in rotating molecules \cite{nonspr}.

The angular momentum production for compact binaries in the final stages of their merger is huge. In the simplest case of two equal masses $m$ in circular orbits with the radius $R$, the angular momentum luminosity is \cite{peters}:
\begin{align}\label{lumin}
\frac{dL}{dt}=\frac{4}{5}\frac{mc^2}{(R/R_S)^{7/2}},
\end{align}
where $R_S$ is the Schwarzschild radius for mass $m$. For example, for two black holes having 30 solar masses inspiraling at a distance of 10 $R_S$ the gravitational radiation carries away every second about six million times the angular momentum of the Earth in its orbital motion.

In contrast to the electromagnetic case, the trapping in the gravitational case is universal; it does not depend on the mass of the body. The simplest model of a gravitational wave carrying angular momentum is the Bessel beam. Bessel beams are non-diffractive; they have the same width along the direction of propagation which is usually chosen as the direction of the $z$ axis. The Riemann tensor for Bessel beams in the weak field approximation has been obtained in \cite{bessel0}. Its components can be constructed \cite{pr} from the components of the rank-four symmetric spinor $\phi_{ABCD}$:
\begin{align}\label{rt}
R_{\mu\nu\alpha\beta}(\bm r,t)=\Re\left( S_{\mu\nu}^{\;\;AB}S_{\alpha\beta}^{\;\;CD}\phi_{ABCD}(\bm r,t)\right),
\end{align}
where $S_{\mu\nu}^{\;\;AB}$ are built from Pauli matrices,
\begin{subequations}
\begin{align}\label{ss}
\{S_{01},S_{02},S_{03}\}&=-i\sigma_y\{\sigma_x,\sigma_y,\sigma_z\},\\
\{S_{23},S_{31},S_{12}\}&=i\{S_{01},S_{02},S_{03}\}.
\end{align}
\end{subequations}
The sign of the Riemann tensor in (\ref{rt}) is opposite to that used in \cite{bessel0} and \cite{pr} because we use now a more common sign convention for the Riemann tensor, instead of the one used in the spinorial formalism.

The five independent components of $\phi_{ABCD}$ for the gravitational Bessel beam can be written in the form \cite{bessel0}:
\begin{align}\label{bes}
\left[\begin{array}{c}
\phi_{0000}\\
\phi_{0001}\\
\phi_{0011}\\
\phi_{0111}\\
\phi_{1111}
\end{array}\right]=A_r\omega^2e^{i\Phi}
\left[\begin{array}{c}
e^{-2i\varphi}J_{M-2}(\kappa\rho)\\
i\lambda q\,e^{-i\varphi}J_{M-1}(\kappa\rho)\\
-q^2J_{M}(\kappa\rho)\\
-i\lambda q^3e^{i\varphi}J_{M+1}(\kappa\rho)\\
q^4e^{2i\varphi}J_{M+2}(\kappa\rho)
\end{array}\right],
\end{align}
where $\Phi=\lambda(k_z z-\omega t+M\varphi)$, $\lambda=\pm 1$ is the beam helicity, $\omega=c\sqrt{k_z^2+\kappa^2}$, $\kappa=\sqrt{k_x^2+k_y^2}$, $\hbar M$ is the component of the {\em total} angular momentum along the beam axis (per one graviton) and $q=c\kappa/\left(\omega+ck_z\right)$. The dimensionless amplitude $A_r$ of the Riemann tensor will be adjusted later. We associated angular momentum with gravitons to make the connection with electromagnetism, where the language of photons is used to describe light beams carrying angular momentum. However, the number of gravitons (if they really exist) is so huge that purely classical analysis is fully justified.

The spinorial components (\ref{bes}) are obtained with the use of the procedure introduced by Penrose \cite{rp} who has shown that every solution of the wave equation for a massless particle can be written (up to a Lorentz transformation) in the form \cite{bessel0}:
\begin{align}\label{sol0}
\phi_{AB\dots L}(\bm r,t)=D_AD_B\dots D_L\,\chi(\bm r,t),
\end{align}
where $\{D_0,D_1\}=\{1/c\,\partial_t-\partial_z,-\partial_x-i\partial_y\}$ and $\chi(\bm r,t)$ is the generating function: a complex solution of the scalar d'Alembert equation. To obtain a gravitational Bessel beam, we choose this solution in the form:
\begin{align}\label{bes0}
\chi_M(\rho,\varphi,z,t)=e^{i\lambda(k_z z-\omega t+M\varphi)}J_M(\kappa\rho).
\end{align}
Exact Bessel beams (both electromagnetic and gravitational) are unphysical, since their energy flux is infinite. However they can be treated as a useful approximation of the radiation from the physical source, when we consider only the region to the vicinity of the beam center.

\begin{figure}
\includegraphics[width=0.3\textwidth]{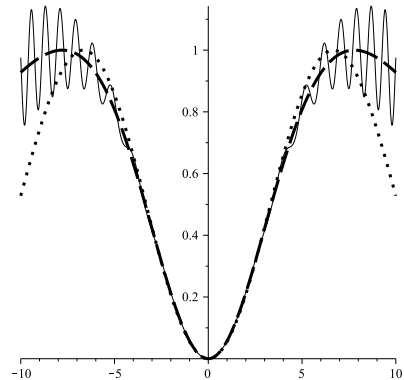}
\vspace{0.6cm}
\caption{The distribution of the $z$-component of angular momentum at 10 wavelengths from the binary system source (solid line) at some given time. Dashed line represents the time averaged distribution. Dotted line represents the Bessel beam  $z$-component angular momentum distribution. Horizontal axis is in units of the wavelength.}\label{fig2}
\end{figure}

The most natural source of gravitational radiation is a close binary system of compact massive objects. The distribution of the angular momentum of the gravitational radiation generated by such system in the linearized approximation can be computed using the framework developed in \cite{butcher1,butcher2}. Consider a binary system moving on quasi-Keplerian orbits in the $xy$-plane. We compute the radiative (time dependent) components of the metric generated by such system using the formula (73) in \cite{butcher1}. Next, we compute the energy-momentum tensor $\tau_{\mu\nu}$ for this metric using the formula (23) from \cite{butcher1}.
Finally the distribution of the angular momentum ${j_{\mu\nu}}^\alpha=2{x_{[\mu}\tau_{\nu]}}^\alpha$ of the gravitational radiation generated by the binary system can be computed using the formula (19) from \cite{butcher2}.
In Fig.~\ref{fig2} we show the distribution of the the $z$-component  of orbital angular momentum (namely ${j_{12}}^0$) in the vicinity of the rotation axis of the system at the distance of 10 wavelengths from the source. It is compared to the distribution of the $z$-component of the orbital angular momentum of the Bessel beam with the angular frequency $\omega=1\,\mathrm{Hz}$, which corresponds to the binary system 4 days before the merger, each component having 30 solar masses and orbiting with angular velocity $\Omega=\omega/2$.

Comparing these two plots we see, that the distributions of orbital angular momentum density in the vicinity of the axis is very similar for both beams. They have minima at the axis. The amplitude of Bessel beam is such that the distributions have the same values at first maximum (time-averaged in case of binary system radiation). The width of the minimum of Bessel beam is controlled by $k_z/\kappa$ ratio. The value $k_z/\kappa=22.7$ for figure  \ref{fig2} was found by minimizing the integral of the square of the difference of the two distributions on the interval around axis in which the time-averaged binary system angular momentum distribution has values smaller than half of first maximum value. We conclude that the description of the gravitational wave in the vicinity
of the wave axis by the Bessel beam is a reasonable approximation.

In this Letter we will only give the proof of concept deferring a detailed analysis to a separate publication. Namely, we will consider the motion of trapped bodies close to the beam axis which enables us to obtain very simple analytic solutions.

Numerical relativity \cite{lsh,lion} leads to the conclusion that the dominant contribution to the gravitational radiation from the binary system comes from $M=2$. In what follows we shall consider only this case. For $M=2$ there is a simple geodesic motion: along the $z$-axis with arbitrary constant velocity. This property can by established by calculating the correction to the flat metric (in the TT gauge) in terms of the Riemann tensor. For monochromatic gravitational waves, such as Bessel beams, the formula reads (cf. \cite{mtw} p. 948):
\begin{align}\label{mtw}
h_{ij}^{TT}=\frac{2}{\omega^2}R_{i0j0}.
\end{align}
The amplitude $A_h$ of the oscillations of the metric for the binary system is given by the formula (23.46) in \cite{jh},
\begin{align}\label{hartle}
A_h=\frac{8G\Omega^2mR^2}{c^4r}=\frac{R_s^2}{2Rr},
\end{align}
where $R$ is the radius of the Kepler orbit and $r$ is the distance to the observation point. We have expressed the amplitude $A_h$ in terms of the black hole Schwarzschild radius $R_s=2GM/c^2$ to exhibit the universality of this formula; when distances are measured in $R_s$, the mass of the object does not matter. Having determined the value of the amplitude $A_h$ we can use (\ref{mtw}) to find that $A_r=A_h/2$.

The motion of the bodies will be found from the equations of the geodesic deviation \cite{mtw,dinv}.
\begin{align}\label{deviation}
\frac{d^2\xi^\mu}{d\tau^2}={R^\mu}_{\alpha\beta\nu}u^\alpha u^\beta \xi^\nu,
\end{align}
where $u^\mu=dx^\mu/d\tau$ is the four-velocity of motion on the reference geodesic, $\xi^\mu$ denotes small departures from the reference geodesic and $\tau$ is the proper time. The reference geodesic will be chosen along the $z$-axis. This is indeed a geodesic line because the Christoffel symbols $\Gamma^\mu_{\alpha\beta}$ obtained from the metric (\ref{mtw}) vanish at $z$-axis for $M=2$ when $\alpha$ and $\beta$ take on the values of 0 or 3. Therefore, the geodesic motion along the $z$-axis is uniform: $d^2x^\mu/d\tau^2=0$. The covariant derivatives in (\ref{deviation}) were replaced by plain derivatives because the relevant components of the Christoffel symbol vanish on the reference geodesic. We will denote by $v$ the velocity of motion along the reference geodesic (c=1),
\begin{align}\label{frame}
u^\alpha=(1,0,0,v)/\sqrt{1-v^2}.
\end{align}
In the equation of the geodesic deviation the Riemann tensor is evaluated on the reference geodesic where most of its components vanish since only $\phi_{0000}$ in the formulas (\ref{bes}) survives. The only nonvanishing terms appearing in the equation of the geodesic deviation (\ref{deviation}) are:
\begin{subequations}\label{riem}
\begin{align}
R^1_{\;\;\alpha\beta\,1}u^\alpha u^\beta&=-A_r\omega^2\cos(\omega t-k_z z),\\
R^1_{\;\;\alpha\beta\,2}u^\alpha u^\beta&=-A_r\omega^2\sin(\omega t-k_z z),\\
R^2_{\;\;\alpha\beta\,1}u^\alpha u^\beta&=-A_r\omega^2\sin(\omega t-k_z z),\\
R^2_{\;\;\alpha\beta\,2}u^\alpha u^\beta&=A_r\omega^2\cos(\omega t-k_z z).
\end{align}
\end{subequations}
Since the components $R^0_{\;\;\alpha\beta\,i}u^\alpha u^\beta$  and $R^3_{\;\;\alpha\beta\,i}u^\alpha u^\beta$ vanish, the motion along the $z$-axis is not influenced by the gravitational wave. The dependence of $\xi^0$ and $\xi^3$ on proper time is linear. In order to secure the applicability of the equations of the geodesic deviation, we must choose the corresponding velocity to be the same as that for the reference geodesic, i.e. $\xi^0=u^0\tau$ and $\xi^3=u^3\tau$. Otherwise the separation $\xi$ would not remain small. The argument of trigonometric functions in (\ref{riem}) becomes then $\omega t-k_z z={\tilde\omega}\tau$, where ${\tilde\omega}=(\omega-v k_z)/\sqrt{1-v^2}$ is the Doppler-shifted wave frequency.

Taking into account the expressions (\ref{riem}), we can rewrite the equations of the geodesic deviation for the two remaining components of $\xi^1=\xi$ and $\xi^2=\eta$ in the form:
\begin{subequations}\label{eqs}
\begin{align}
\frac{d^2\xi}{d\tau^2}&=-\gamma{\tilde\omega}^2(\xi\cos({\tilde\omega} \tau)+\eta\sin({\tilde\omega}\tau)),\\
\frac{d^2\eta}{d\tau^2}&=-\gamma{\tilde\omega}^2(\xi\sin({\tilde\omega} \tau)-\eta\cos({\tilde\omega}\tau)),
\end{align}
\end{subequations}
where $\gamma=(\omega/{\tilde\omega})^2A_r$. The explicit dependence on the proper time in (\ref{eqs}) can be eliminated by the following transformation to the rotating frame,
\begin{subequations}\label{rot}
\begin{align}
\xi(\tau)&=x(\tau)\cos({\tilde\omega}\tau/2)-y(\tau)\sin({\tilde\omega} \tau/2),\\
\eta(\tau)&=x(\tau)\sin({\tilde\omega}\tau/2)+y(\tau)\cos({\tilde\omega} \tau/2).
\end{align}
\end{subequations}
The equations of motion in the rotating frame are:
\begin{subequations}\label{eqs1}
\begin{align}
\ddot{x}(\tau)&=(1/4-\gamma){\tilde\omega}^2 x(\tau)+{\tilde\omega} \dot{y}(\tau),\\
\ddot{y}(\tau)&=(1/4+\gamma){\tilde\omega}^2 y(\tau)-{\tilde\omega} \dot{x}(\tau).
\end{align}
\end{subequations}
The solutions of these linear equations with constant coefficients are obtained by the standard substitution,
\begin{align}
x(\tau)=a e^{i\lambda\tau},\quad y(\tau)=b e^{i\lambda\tau}.
\end{align}
The nonzero solutions exist when the determinant of the set of two equations,
\begin{align}
\lambda^4-{\tilde\omega}^2\lambda^2/2+(1/16-\gamma^2){\tilde\omega}^4,
\end{align}
vanishes. The four characteristic frequencies are: $\lambda^\pm_\pm=\pm\tilde\omega\sqrt{1/4\pm\gamma}$. The general solution of (\ref{eqs1}) is a linear combination of the solutions containing all four frequencies with the coefficients depending on the initial data. These solutions have the same general properties as those found in Refs.\cite{danby,paul,lagr,nonspr}. Namely, due to the Coriolis force, the test particles will be trapped by the gravitational wave, they oscillate near the wave center.

After the transformation back to the laboratory frame, with the use of the formulas (\ref{rot}), we obtain the following final expression for the complex combination $\zeta(\tau)=\xi(\tau)+i\eta(\tau)$:
\begin{align}\label{sol}
\zeta(\tau)=e^{i\sigma}[c_{++}e^{i\sigma\sqrt{1+4\gamma}}
+c_{+-}e^{-i\sigma\sqrt{1+4\gamma}}\nonumber\\
+c_{-+}e^{i\sigma\sqrt{1-4\gamma}}
+c_{--}e^{-i\sigma\sqrt{1-4\gamma}}],
\end{align}
where $\sigma={\tilde\omega}\tau/2$ and
\begin{subequations}\label{cc}
\begin{align}
&c_{+\pm}=(\mp q_1+r_1)/2,\quad c_{-\pm}=-i(\pm q2-r2)/2,\\
&q_1=\frac{\xi_0+i(1+4\gamma)\xi_R-\eta_R}
{\sqrt{1+4\gamma}},\;r_1=\xi_0+i\xi_R-\eta_R,\\
&q_2=\frac{\eta_0-i(1-4\gamma)\eta_R-\xi_R}
{\sqrt{1-4\gamma}},\;r_2=\eta_0-i\eta_R-\xi_R.
\end{align}
\end{subequations}
The initial positions are $\xi(0)=\xi_0,\;\eta(0)=\eta_0$ and the initial velocities $\xi'(0)$ and $\eta'(0)$ appear in these formulas always in the renormalized form $\xi_R=\xi'(0)/(2\gamma{\tilde\omega}),\;
\eta_R=\eta'(0)/(2\gamma{\tilde\omega})$.

Stable solutions exist only when the gravitational wave is not too strong: $|\gamma|< 1/4$. However, such an extreme value of $\gamma$ is totally unrealistic. It corresponds to both $R$ and $r$ equal to the Schwarzschild radius. Of course, in the limit when the gravitational wave is turned off completely ($\gamma=0$) the motion is uniform with the initial velocity, $\xi(\tau)=\xi_0+\xi'_0\tau,\;\eta(\tau)
=\eta_0+\eta'_0\tau$.

\begin{figure}
\includegraphics[width=0.185\textwidth]{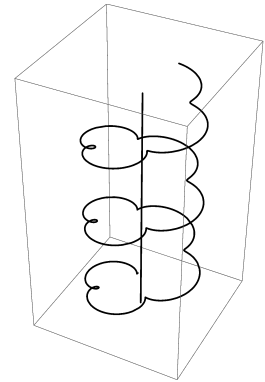}
\caption{Reference geodesic (straight line along the $z$-axis) and a nearby geodesic guided by the gravitational wave. The units are not shown in this figure because the size of the curves is arbitrary; it depends solely on the initial conditions. The initial conditions $\xi_0=\eta_R$ and $\gamma=5/36$ are chosen to obtain a strictly periodic motion.} \label{fig3}
\end{figure}

A trapping of a body by the gravitational Bessel wave is shown in Fig.\ref{fig3}. The periodic geodesic winding around the reference geodesic proves that trapping is permanent, The stability of the trapping for all values of parameters is best illustrated by the average distance $D$ from the beam center,
\begin{align}\label{dist}
D^2&=\frac{(1+2\gamma)(\xi_0-\eta_R)^2}{1+4\gamma}
+\frac{(1-2\gamma)(\eta_0-\xi_R)^2}{1-4\gamma}\nonumber\\
&+(1+2\gamma)\xi_R^2+(1-2\gamma)\eta_R^2.
\end{align}
In our analysis we disregarded the gravitational pull exerted by the rotating binary system because it is predominantly directed along the $z$-asis. Therefore, it does not affect the trapping in the $xy$-plane.

To prove that the role of angular momentum is absolutely essential for trapping, we derived for $M=0$ the counterpart of equations (\ref{eqs}):
\begin{subequations}\label{eqs0}
\begin{align}
\frac{d^2\xi}{d\tau^2}&=a\,\xi\cos(\omega\tau)+b\,\eta\sin(\omega\tau)),\\
\frac{d^2\eta}{d\tau^2}&=-b\,\xi\sin(\omega\tau)+a\,\eta\cos(\omega\tau)),
\end{align}
\end{subequations}
where $a$ and $b$ are real coefficients built from the wave parameters. These equations have only runaway solutions in the form of Matthieu functions of complex arguments.

We have chosen a simple radiating system: two orbiting compact objects. In this case it was possible to give in the paraxial approximation a complete analytic solution of the trapping phenomenon. The trapping by the gravitational wave carrying angular momentum, however, is universal. It can also take place at the galactic scale where it may play a role in the formation of galactic jets.
\vspace{0.2cm}

Note added in proof: After the submission of this Letter the paper has appeared (Phys. Rev. D {\bf 98}, 044037 (2018)) and the manuscript by the same authors was posted (arXiv:1807.00765v3 [gr-qc]), where the connection between the electromagnetic trapping (Paul trap) and the trapping by gravitational waves has been discussed from a totally different perspective.

\end{document}